\documentclass[a4paper,11pt]{article}
\usepackage{amsfonts,slashed}
\usepackage{url}
\usepackage{latexsym}
\usepackage{amsmath}
\usepackage{amssymb}
\usepackage{indentfirst}
\usepackage{graphicx}
\usepackage{epsfig}
\usepackage{amsthm}
\usepackage{amsfonts}
\usepackage{indentfirst}
\usepackage[colorlinks]{hyperref}
\usepackage{cite}
\usepackage{cancel}
\usepackage{tikz}
\usepackage{enumitem}

\begin{document}

\begin{titlepage}


\title{\bf{An action principle for the Einstein-Weyl equations}} 
\author{Silke Klemm$^{1,2}$\thanks{silke.klemm@mi.infn.it} and Lucrezia Ravera$^{3,4}$\thanks{lucrezia.ravera@polito.it} \\ \\
{\small $^{1}$\textit{Dipartimento di Fisica, Universit\`{a} di Milano, Via Celoria 16, 20133 Milano, Italy}}\\
{\small $^{2}$\textit{INFN, Sezione di Milano, Via Celoria 16, I-20133 Milano, Italy}}\\
{\small $^{3}$\textit{DISAT, Politecnico di Torino, Corso Duca degli Abruzzi 24, 10129 Torino, Italy}}\\
{\small $^{4}$\textit{INFN, Sezione di Torino, Via P. Giuria 1, 10125 Torino, Italy}}}
\clearpage\maketitle
\thispagestyle{empty}

\begin{abstract}
A longstanding open problem in mathematical physics has been that of finding an action principle for
the Einstein-Weyl (EW) equations.
In this paper, we present for the first time such an action principle in three dimensions in which the Weyl vector is not exact. More precisely, our model contains, in addition to
the Weyl nonmetricity, a traceless part. If the latter is (consistently) set to zero, the equations of motion
boil down to the EW equations.
In particular, we consider a metric affine $f(R)$ gravity
action plus additional terms involving Lagrange multipliers and gravitational Chern-Simons contributions.
In our framework, the metric and the connection are considered as independent objects, and no a priori assumptions on the nonmetricity and the torsion of the connection are made. The dynamics of the
Weyl vector turns out to be governed by a special case of the generalized monopole equation, which 
represents a conformal self-duality condition in three dimensions.
\end{abstract}

\vspace{2cm}


\end{titlepage}

\section{Introduction}

A remarkable generalization of Riemannian geometry was first proposed in 1918 by
Weyl (see e.g. \cite{Adler:1965,CP1,Folland,Romero:2012hs}), who introduced an additional
symmetry in an attempt of geometrically
unifying electromagnetism with gravity \cite{Weyl:1918ib, Weyl:1919fi}. In this theory, both the direction and the length of vectors are allowed to vary under parallel transport\footnote{Note, however, that Weyl's
attempt to identify the trace part of the connection, associated with stretching and contraction, with the 
vector potential of electromagnetism failed, due to observational inconsistencies \cite{Wheeler:2018rjb}.}.
The trace part of the connection introduced by Weyl is known as the Weyl vector. When it is exact,
it can be gauged away by a local scale transformation. In this case, Weyl geometry is said to be integrable
(parallel transported vectors along closed paths return with unaltered lengths), and there exists a subclass
of global gauges in which the geometry is Riemannian.

Mathematically, a Weyl structure on an $n$-dimensional manifold $\cal M$ consists of a conformal
structure $[g]=\{fg,f:{\cal M}\to\mathbb{R}^+\}$, together with a torsion-free connection $\nabla$
which is compatible with $[g]$ in the sense that
\begin{equation}
\nabla_\mu g_{\nu\rho} = 2 h_{\mu} g_{\nu\rho}\,, \label{eq:Weyl-nonmetr}
\end{equation}
for some one-form $h$ on $\cal M$. This compatibility condition is invariant under the transformation
\begin{equation}
g\mapsto e^{2\Omega} g\,, \qquad h\mapsto h + d\Omega\,,
\end{equation}
where $\Omega$ is a function on $\cal M$.

A Weyl structure is said to be Einstein-Weyl (EW) \cite{Hitchin:1982vry} if the symmetrized
Ricci tensor $R_{\mu\nu}$ of $\nabla$ is proportional to some metric $g\in [g]$. This conformally
invariant condition is equivalent to
\begin{equation}
R_{(\mu\nu)} - \frac Rn g_{\mu\nu} = 0\,, \label{EW-eqns}
\end{equation}
with $R$ the Ricci scalar of $\nabla$.

A longstanding and still unresolved open problem has been that of finding an action principle
for \eqref{EW-eqns}. Here we shall present for the first time an action principle for the EW equations in
$2+1$ dimensions\footnote{Our results hold also in three Euclidean dimensions.}
in which the Weyl vector is not exact. Notice that three-dimensional EW geometry \cite{pt}
is particularly interesting, since it is related to dispersionless integrable
systems \cite{Ward:1990qt,Dunajski:2000rf,Dunajski:2008xx}.
Moreover, Jones and Tod \cite{Jones:1985pla} showed that selfdual conformal four-manifolds with
a conformal vector field are in correspondence with abelian monopoles on Einstein-Weyl three-manifolds.
In ref.~\cite{Gauduchon:1998xx}, Gauduchon and Tod studied the structure of four-dimensional
hyper-Hermitian Riemannian spaces admitting a tri-holomorphic Killing vector, i.e., a Killing vector
that is compatible with the three complex structures on the hyper-Hermitian space. It turned out that
the latter is fibered over a specific type of three-dimensional EW spaces, called hyper-CR or
Gauduchon-Tod.

Note also that 3d EW manifolds and their generalizations play an important role also in high energy
physics, for instance in the context of supersymmetric solutions to fake
supergavity \cite{Meessen:2009ma,Gutowski:2009vb,Grover:2009ms}, in the classification
of four-dimensional Euclidean gravitational instantons \cite{Dunajski:2010zp,Dunajski:2010uv},
or supersymmetric near-horizon geometries \cite{Dunajski:2016rtx, Klemm:2019izb}. Moreover, Weyl connections were considered recently in holography \cite{Ciambelli:2019bzz}. For a general review of
Einstein manifolds with nonmetric and torsionful connections cf.~\cite{Klemm:2018bil}.

Our construction of an action that leads to \eqref{EW-eqns} involves a metric affine $f(R)$ gravity
action \cite{Hehl:1994ue, Iosifidis:2018zjj, Iosifidis:2019jgi, Sotiriou:2006qn, Sotiriou:2006mu, Vitagliano:2010sr} plus additional terms containing Lagrange multipliers and gravitational Chern-Simons 
contributions. We work in a first order formalism, where the metric and the connection are treated as
independent variables, and make no a priori assumptions on the metricity and the torsion
of the connection.

The remainder of this paper is organized as follows: In section \ref{review} we briefly review
metric affine $f(R)$ gravity in $n$ dimensions, following \cite{Iosifidis:2018zjj,Iosifidis:2019jgi}. Subsequently, in section \ref{magin3dplusCS} we construct an action that yields the EW equations
with nonexact Weyl vector in three dimensions. We conclude our work with some final remarks.

\section{Metric affine $f(R)$ gravity in $n$ dimensions}\label{review}

Consider the gravitational action
\begin{equation}
S_{\text{G}} = \frac1{2\kappa^2}\int d^n x\sqrt{-g} f(R)\,, \label{action-f(R)}
\end{equation}
where $\kappa$ denotes the gravitational coupling constant and $f(R)$ is an arbitrary function
of the scalar curvature $R = g^{\mu \nu} R_{\mu \nu}(\Gamma)$, with $\Gamma$ a general affine 
connection\footnote{We adopt the same conventions of \cite{Iosifidis:2018zjj,Iosifidis:2019jgi}. In
particular, our metric convention is $\eta= \text{diag} (-, +, +)$.}.
We work in a first order (Palatini) formalism, where the metric $g_{\mu\nu}$ and the connection 
${\Gamma^\lambda}_{\mu\nu}$ are treated as independent variables. Variation of \eqref{action-f(R)}
w.r.t.~$g^{\mu\nu}$ gives
\begin{equation}
f'(R) R_{(\mu\nu)} - \frac12 f(R) g_{\mu\nu} = 0\,, \label{eq:Einstein}
\end{equation}
while the variation w.r.t.~${\Gamma^\lambda}_{\mu\nu}$ leads to
\begin{equation}\label{eq:Conn}
\begin{split}
& \frac1{\sqrt{-g}}\left[-\nabla_\lambda\left(\sqrt{-g} f'(R) g^{\mu\nu}\right) + \nabla_\sigma\left(\sqrt{-g} f'(R) g^{\mu\sigma}\right){\delta^\nu}_\lambda\right] \\
& + 2f'(R)\left(g^{\mu\nu}{\Gamma^\sigma}_{[\lambda\sigma]} - g^{\mu\rho}
{\Gamma^\sigma}_{[\rho\sigma]}{\delta^\nu}_\lambda + g^{\mu\sigma}{\Gamma^\nu}_{[\sigma\lambda]}
\right) = 0\,.
\end{split}
\end{equation}
The trace of \eqref{eq:Einstein} yields
\begin{equation}\label{trace}
\frac{f}{2f'} = \frac Rn\,,
\end{equation}
which is identically satisfied if we choose
\begin{equation}
f = CR^{n/2}\,, \label{eq:f}
\end{equation}
where $C$ is an arbitrary integration constant. Let us observe that \eqref{trace} could also be viewed
as an algebraic equation on $R$ admitting generically solutions with constant scalar curvature
(cf.~e.g.~\cite{Iosifidis:2019jgi}). Here, we consider the specific choice \eqref{eq:f}. Then (with $C=1$),
the action \eqref{action-f(R)} becomes
\begin{equation}
S_{\text{G}} = \frac1{2\kappa^2}\int d^n x \sqrt{-g} R^{n/2} \,, \label{action-f(R)-fixed}
\end{equation}
which is invariant under the conformal transformation (we use the same definition as \cite{Iosifidis:2019jgi})
\begin{equation}\label{weylresc}
g_{\mu\nu}\mapsto g'_{\mu\nu} = e^{2\Omega} g_{\mu\nu}\,, \qquad
{\Gamma^\lambda}_{\mu\nu}\mapsto {\Gamma'^\lambda}_{\mu\nu} = {\Gamma^\lambda}_{\mu\nu}\,,
\end{equation}
where $\Omega$ is a scalar function. Indeed, under \eqref{weylresc} we have
\begin{equation}\label{conftr}
d^n x\sqrt{-g}\mapsto d^n x e^{n\Omega}\sqrt{-g}\,, \qquad
{R^\lambda}_{\mu\nu\rho}\mapsto {R^\lambda}_{\mu\nu\rho}\,, \qquad
R_{\mu\rho}\mapsto R_{\mu\rho}\,, \qquad R\mapsto e^{-2\Omega} R\,,
\end{equation}
and one can clearly see that \eqref{action-f(R)-fixed} is invariant.

Now, plugging \eqref{trace} into \eqref{eq:Einstein}, the latter boils down to
\begin{equation}
R_{(\mu\nu)} - \frac Rn g_{\mu\nu} = 0\,. \label{eq:EW}
\end{equation}
In the case in which one has Weyl nonmetricity \eqref{eq:Weyl-nonmetr} and vanishing torsion,
\eqref{eq:EW} precisely corresponds to the Einstein-Weyl equations. Observe that \eqref{eq:EW} is traceless.

Let us take a closer look at \eqref{eq:Conn}. Using the definition of the Cartan torsion tensor,
\begin{equation}
{T_{\mu\nu}}^\rho := {\Gamma^\rho}_{[\mu\nu]}\,, \label{eq:torsdef}
\end{equation}
\eqref{eq:Conn} can be rewritten as
\begin{equation}\label{eq:ConnTor}
\begin{split}
& \frac1{\sqrt{-g}}\left[-\nabla_\lambda\left(\sqrt{-g} f'(R) g^{\mu\nu}\right) + \nabla_\sigma\left(\sqrt{-g}f'(R) g^{\mu\sigma}\right){\delta^\nu}_\lambda\right] \\
& + 2f'(R)\left(g^{\mu\nu}T_\lambda - T^\mu{\delta^\nu}_\lambda + g^{\mu\sigma}{T_{\sigma \lambda}}^{\nu}
\right) = 0\,,
\end{split}
\end{equation}
where $T_\lambda :={T_{\lambda \sigma}}^{\sigma}$ is the trace part of the torsion.
Taking the $\lambda,\mu$ trace of \eqref{eq:ConnTor} leads to the identity
\begin{equation}\label{palatinitraceless}
{P_{\mu}}^{\mu \nu} = 0\,,
\end{equation}
where we introduced the so-called Palatini tensor,
\begin{equation}\label{palatini}
{P_{\lambda}}^{\mu\nu} = -\frac{\nabla_\lambda\left(\sqrt{-g} g^{\mu\nu}\right)}{\sqrt{-g}} + \frac{\nabla_\sigma\left(\sqrt{-g}g^{\mu\sigma}\right){\delta^\nu}_\lambda}{\sqrt{-g}} + 2 \left(g^{\mu\nu}T_\lambda - T^\mu{\delta^\nu}_\lambda + g^{\mu\sigma}{T_{\sigma\lambda}}^\nu
\right)\,,
\end{equation}
which is indeed traceless.
Contracting $\lambda$ and $\nu$ in \eqref{eq:ConnTor}, one obtains\footnote{Here we also fix a 
miscalculation appearing in \cite{Sotiriou:2006qn}.}
\begin{equation}
(n-1)\nabla_\sigma\left(\sqrt{-g} f'(R) g^{\sigma\mu}\right) = 2(n-2)\sqrt{-g} f'(R) T^\mu\,.
\label{eq:contractionln}
\end{equation}
Plugging \eqref{eq:contractionln} into \eqref{eq:ConnTor} and taking the symmetric and antisymmetric
part in $\mu,\nu$, we get respectively
\begin{equation}
- \frac1{\sqrt{-g}}\nabla_\lambda \left(\sqrt{-g} f'(R) g^{\mu\nu}\right)+ 2f'(R)\left(g^{\mu\nu}T_\lambda - \frac{1}{n-1} g^{\rho(\mu}{\delta^{\nu)}}_\lambda T_\rho + g^{\sigma(\mu}{T_{\sigma\lambda}}^{\nu)}
\right) = 0\,,
\label{eq:Conn-symm}
\end{equation}
and
\begin{equation}
-\frac1{n-1} g^{\rho[\mu}{\delta^{\nu]}}_\lambda T_\rho + g^{\sigma[\mu}
{T_{\sigma\lambda}}^{\nu]} = 0\,.
\label{eq:Conn-antisymm}
\end{equation}
After some algebraic manipulation, \eqref{eq:Conn-antisymm} gives
\begin{equation}\label{torsionobtained}
{T_{\lambda \mu}}^{\nu} = \frac{2}{n-1} T_{[\lambda} {\delta_{\mu]}}^{\nu}\,,
\end{equation}
i.e., the torsion is completely determined by its trace part $T_\lambda$.
On the other hand, if we take the $\mu,\nu$ trace in \eqref{eq:Conn-symm} and use the general
formula \cite{Iosifidis:2018zjj,Iosifidis:2019jgi}
\begin{equation}
\nabla_\lambda\sqrt{-g} = -\frac12\sqrt{-g} g_{\mu\nu}\nabla_\lambda g^{\mu\nu} =
-\frac12 \sqrt{-g} g_{\mu\nu} {Q_\lambda}^{\mu\nu}= - \frac12\sqrt{-g} Q_\lambda\,,
\end{equation}
where $Q_\lambda \equiv {Q_{\lambda \mu}}^\mu$ is a trace of the nonmetricity
tensor\footnote{The latter is defined by $Q_{\lambda\mu\nu}:= -\nabla_\lambda g_{\mu\nu} = 
-\partial_\lambda g_{\mu\nu} + {\Gamma^\rho}_{\mu\lambda} g_{\rho\nu} +
{\Gamma^\rho}_{\nu\lambda}g_{\mu\rho}$.}, we find
\begin{equation}\label{QT=df'/f'}
\frac{n-2}{2n} Q_\lambda + \frac{2(n-2)}{n-1} T_\lambda = \frac{\partial_\lambda f'}{f'}\,,
\end{equation}
or equivalently
\begin{equation}\label{affinevector=df'/f'}
\frac{n-2}2\mathit{w}_\lambda = \partial_\lambda \text{ln} f'\,,
\end{equation}
where we introduced the so-called affine vector $\mathit{w}_\lambda$ \cite{Iosifidis:2018zjj}, defined in
$n$ dimensions by
\begin{equation}\label{affinevector}
\mathit{w}_\lambda :=\frac1n Q_\lambda + \frac4{n-1} T_\lambda\,.
\end{equation}
Observe that, using \eqref{eq:f}, eq.~\eqref{affinevector=df'/f'} can be rewritten as
\begin{equation}\label{dlnRandaffvector}
\partial_\lambda\text{ln} R = \mathit{w}_\lambda\,,
\end{equation}
i.e., the affine vector is exact.
Now, plugging the expression for $\partial_\lambda\ln f'$ from \eqref{QT=df'/f'} and the torsion
\eqref{torsionobtained} into \eqref{eq:Conn-symm}, we obtain
\begin{equation}\label{nonmetrobtained}
Q_{\lambda\mu\nu} = \frac1n Q_\lambda g_{\mu\nu}\,.
\end{equation}
The nonmetricity is thus fully determined by the vector $Q_\lambda$. Notice that in the irreducible
decomposition of the nonmetricity tensor under the Lorentz group there is a second nonmetricity vector
$\tilde{Q}_\nu := {Q^\mu}_{\mu\nu}$. Using \eqref{nonmetrobtained}, one gets
$\tilde{Q}_\nu = \frac1n Q_\nu$, and therefore \eqref{nonmetrobtained} can alternatively be written as
$Q_{\lambda\mu\nu} = \tilde{Q}_\lambda g_{\mu\nu}$.

Finally, exploiting the generic decomposition of an affine connection,
\begin{equation}\label{gendecompaffconn}
{\Gamma^\lambda}_{\mu\nu} = \tilde{\Gamma}^\lambda_{\phantom{\lambda}\mu\nu} + {N^\lambda}_{\mu\nu}\,,
\end{equation}
where the distortion tensor ${N^\lambda}_{\mu\nu}$ and the Levi-Civita connection 
$\tilde{\Gamma}^\lambda_{\phantom{\lambda}\mu\nu}$ are respectively given by
\begin{equation}\label{distortion}
{N^\lambda}_{\mu\nu} = \underbrace{\frac12 g^{\rho\lambda}\left(Q_{\mu\nu\rho} + Q_{\nu\rho\mu}
- Q_{\rho\mu\nu}\right)}_{\text{deflection}} - \underbrace{g^{\rho\lambda}\left(T_{\rho\mu\nu} +
T_{\rho\nu\mu} - T_{\mu\nu\rho}\right)}_{\text{contorsion}}\,,
\end{equation}
\begin{equation}
\tilde{\Gamma}^\lambda_{\phantom{\lambda}\mu\nu} = \frac12 g^{\rho\lambda}\left(\partial_\mu 
g_{\nu\rho} + \partial_\nu g_{\rho\mu} - \partial_\rho g_{\mu\nu}\right)\,,
\end{equation}
one can show that, in the present case, the complete expression for the affine connection reads
\begin{equation}\label{affineconnobtained}
{\Gamma^\lambda}_{\mu\nu} = \tilde{\Gamma}^\lambda_{\phantom{\lambda}\mu\nu} + \frac1{2n} Q_\nu
{\delta^\lambda}_\mu + \frac12\left(\mathit{w}_\mu {\delta_\nu}^\lambda - g_{\mu\nu}
\mathit{w}^{\lambda}\right)\,.
\end{equation}
Notice that, in $f(R)$ theories, the connection is only determined up to a vectorial degree of
freedom \cite{Iosifidis:2018zjj,Iosifidis:2019jgi}. Indeed, using the definition of the Ricci tensor,
\begin{equation}
R_{\mu\nu} := {R^\lambda}_{\mu\lambda\nu} = 2\partial_{[\lambda} {\Gamma^\lambda}_{|\mu|\nu]}
+2 {\Gamma^{\lambda}}_{\rho[\lambda} {\Gamma^\rho}_{|\mu|\nu]} \,,
\end{equation}
one can show that under projective transformations\footnote{The latter are defined as those
transformations of the affine connection that leave the autoparallels of vectors invariant up to 
reparametrizations of the affine parameter \cite{Iosifidis:2018zjj}.}
\begin{equation}\label{projectivetransf}
{\Gamma^\lambda}_{\mu\nu}\mapsto {\Gamma'^\lambda}_{\mu\nu} = {\Gamma^\lambda}_{\mu\nu} + 
{\delta^\lambda}_\mu\xi_\nu\,,
\end{equation}
where $\xi_\nu$ is an arbitrary vector field, the symmetric part of the Ricci tensor and thus the Ricci
scalar and the action \eqref{action-f(R)} remain invariant.

Observe that in the case in which \eqref{trace} is considered as a purely algebraic equation for the scalar 
curvature $R$ yielding constant curvature metrics, one has, from \eqref{dlnRandaffvector}, that the affine 
vector $w_\mu$ vanishes, which implies
\begin{equation}
Q_\mu = - \frac{4n}{n-1} T_\mu\,,
\end{equation}
and
\begin{equation}
{\Gamma^\lambda}_{\mu\nu} = \tilde{\Gamma}^\lambda_{\phantom{\lambda}\mu\nu} +
\frac1{2n} Q_\nu {\delta^\lambda}_{\mu}\,.
\end{equation}
One can therefore always arrive at the Levi-Civita connection by choosing $\xi_\nu$ appropriately in
\eqref{projectivetransf}.

On the other hand, in the case of $f(R)$ theories with $f(R) = R^{n/2}$, taking into account that under the 
projective transformation \eqref{projectivetransf} one has\footnote{Here we use
$N_{(\lambda\mu)\nu} = \frac12 Q_{\nu\lambda\mu}$.}
\begin{equation} 
\begin{split}
& {T_{\mu\nu}}^\lambda\mapsto {T_{\mu\nu}}^\lambda + {\delta_{[\mu}}^\lambda\xi_{\nu]}\,, \qquad
T_\mu\mapsto T_\mu - \frac{n-1}2\xi_\mu\,, \\
& {N^\lambda}_{\mu\nu}\mapsto {N^\lambda}_{\mu\nu} + {\delta^\lambda}_{\mu}\xi_\nu\,, \qquad
Q_{\lambda\mu\nu}\mapsto Q_{\lambda\mu\nu} + 2\xi_\lambda g_{\mu\nu}\,, \qquad
Q_\mu\mapsto Q_\mu + 2 n\xi_\mu\,,
\end{split}
\end{equation}
one easily shews that the affine vector $w_\mu$ is invariant under \eqref{projectivetransf}. This is
consistent with the fact that $w_\mu$ is related by \eqref{dlnRandaffvector} to the scalar curvature $R$,
which was shown above to be projectively invariant. One can use the projective gauge freedom
to further restrict the torsion and nonmetricity, and thus the affine connection. For instance, we can
choose $\xi_\mu$ in such a way to eliminate either the torsion or the
nonmetricity \cite{Iosifidis:2018zjj,Iosifidis:2019jgi},
\begin{equation}
\begin{split}
&\xi_\mu = \frac2{n-1} T_\mu\, \Rightarrow\,
\left\{
\begin{aligned}
& \, T'_\mu = 0 \, \Rightarrow\, {T'_{\mu\nu}}^\lambda = 0\,, \\
& \, Q'_\mu = Q_\mu + \frac{4n}{n-1} T_\mu\, \Rightarrow\, Q'_{\lambda\mu\nu} = Q_{\lambda\mu\nu}
+ \frac4{n-1} T_\lambda g_{\mu \nu}\,;
\end{aligned}
\right. \\
& \xi_\mu = -\frac1{2n} Q_\mu\, \Rightarrow\,
\left\{
\begin{aligned}
& \, T'_\mu = T_\mu + \frac{n-1}{4n} Q_\mu\, \Rightarrow\, {T'_{\mu\nu}}^\lambda  =
{T_{\mu\nu}}^\lambda + \frac1{2n} Q_{[\mu} {\delta_{\nu]}}^\lambda\ , \\
& \, Q'_\mu = 0\, \Rightarrow\, Q'_{\lambda\mu\nu} = 0\,.
\end{aligned}
\right.
\end{split}
\end{equation}
Intriguingly, one may also choose $\xi_\mu$ in such a way that the torsion and nonmetricity vectors
result to be interchanged \cite{Iosifidis:2018zjj}\footnote{Notice that, in this case, the scalar product
$T_\mu Q^\mu$ is left invariant under the transformation.},
\begin{equation}
\xi_\mu = \frac2{n-1} T_\mu - \frac1{2n} Q_\mu\, \Rightarrow\,
\left\{
\begin{aligned}
& \, T'_\mu = \frac{n-1}{4n} Q_\mu\, \Rightarrow\, {T'_{\mu\nu}}^\lambda = \frac1{2n}
Q_{[\mu} {\delta_{\nu]}}^\lambda\,, \\
& \, Q'_\mu = \frac{4n}{n-1} T_\mu\, \Rightarrow\, Q'_{\lambda\mu\nu} = \frac4{n-1} T_\lambda
g_{\mu\nu}\,.
\end{aligned}
\right. 
\end{equation}
In this sense, systems with spacetime dislocations (induced by torsion) are physically equivalent to
systems with Weyl nonmetricity \cite{Iosifidis:2018zjj}.

As we already mentioned, the action \eqref{action-f(R)-fixed} is also invariant under the conformal
transformations \eqref{weylresc}, under which the torsion and nonmetricity transform respectively as
\begin{equation}
{T_{\mu\nu}}^\lambda\mapsto {T'_{\mu\nu}}^\lambda = {T_{\mu\nu}}^\lambda\,, \qquad
Q_{\lambda\mu\nu}\mapsto Q'_{\lambda\mu\nu} = e^{2\Omega}\left[Q_{\lambda\mu\nu}
- 2 g_{\mu\nu}\partial_\lambda\Omega\right]\,,
\end{equation}
which implies
\begin{equation}
T_\mu\mapsto T'_\mu = T_\mu\,, \qquad
Q_\mu\mapsto Q'_\mu = Q_\mu - 2 n\partial_\mu\Omega\,, \qquad
\tilde{Q}_\mu\mapsto\tilde{Q}'_\mu = \tilde{Q}_\mu - 2\partial_\mu\Omega\,.
\end{equation}
For further details on transformations in metric affine manifolds we refer to \cite{Iosifidis:2019jgi}.

To conclude this preliminary review, let us now briefly discuss, following \cite{Iosifidis:2019jgi}, what 
happens when one adds a matter action $S_{\text M} [g_{\mu\nu}, {\Gamma^\lambda}_{\mu\nu}]$ to
\eqref{action-f(R)}. The full action is then
\begin{equation}\label{full}
S = S_{\text G} + S_{\text M}\,.
\end{equation}
Variation of \eqref{full} w.r.t.~$g^{\mu\nu}$ gives
\begin{equation}
f'(R) R_{(\mu\nu)} - \frac12 f(R) g_{\mu\nu} = \kappa^2 T_{\mu\nu}\,, \label{eq:Einstein+matter}
\end{equation}
where
\begin{equation}\label{enmomtens}
T_{\mu\nu} := -\frac2{\sqrt{-g}}\frac{\delta S_{\text M}}{\delta g^{\mu\nu}}
\end{equation}
is the energy-momentum tensor\footnote{If $f(R)$ is given by \eqref{eq:f}, the trace
of \eqref{eq:Einstein+matter} yields $g^{\mu\nu} T_{\mu\nu}=0$, so this specific model can be coupled
only to conformally invariant matter.},
while the variation w.r.t.~${\Gamma^\lambda}_{\mu\nu}$ leads to
\begin{equation}\label{eq:Conn+matter}
\begin{split}
& \frac1{\sqrt{-g}}\left[-\nabla_\lambda\left(\sqrt{-g} f'(R) g^{\mu\nu}\right) + \nabla_\sigma
\left(\sqrt{-g} f'(R) g^{\mu\sigma}\right){\delta^\nu}_\lambda\right] \\
& + 2f'(R)\left(g^{\mu\nu}{\Gamma^\sigma}_{[\lambda\sigma]} - g^{\mu\rho}
{\Gamma^\sigma}_{[\rho\sigma]}{\delta^\nu}_\lambda + g^{\mu\sigma}{\Gamma^\nu}_{[\sigma\lambda]}
\right) = \kappa^2 {\Delta_\lambda}^{\mu\nu}\,,
\end{split}
\end{equation}
with the hypermomentum tensor
\begin{equation}
{\Delta_\lambda}^{\mu \nu}\equiv - \frac2{\sqrt{-g}}\frac{\delta S_{\text M}}
{\delta {\Gamma^\lambda}_{\mu\nu}}\,,
\end{equation}
that contains information on the spin, shear and dilation of matter. Using the Cartan torsion tensor,
\eqref{eq:Conn+matter} can be cast into
\begin{equation}\label{eq:ConnTor+matter}
\begin{split}
& \frac1{\sqrt{-g}}\left[-\nabla_\lambda\left(\sqrt{-g} f'(R) g^{\mu\nu}\right) +
\nabla_\sigma\left(\sqrt{-g} f'(R) g^{\mu\sigma}\right){\delta^\nu}_\lambda\right] \\
& + 2f'(R)\left(g^{\mu\nu} T_\lambda - T^\mu{\delta^\nu}_\lambda +
g^{\mu\sigma} {T_{\sigma \lambda}}^\nu\right) = \kappa^2 {\Delta_\lambda}^{\mu\nu}\,.
\end{split}
\end{equation}
The $\lambda,\mu$ trace of \eqref{eq:ConnTor+matter} gives
\begin{equation}\label{condonmatter}
{\Delta_\mu}^{\mu\nu} = 0\,,
\end{equation}
which cannot hold for any form of matter. \eqref{condonmatter} arises due to the projective invariance
of the Ricci scalar. Metric affine $f(R)$ theories of gravity can thus
be consistently coupled only to projectively invariant matter. Note in this context that the terms
added to \eqref{action-f(R)-fixed} in the next section break projective invariance, so that the
resulting theory can be coupled to any type of matter.

\section{An action principle for the Einstein-Weyl equations in three dimensions}\label{magin3dplusCS}

In this section, we present an action principle for the Einstein-Weyl equations in $2+1$ dimensions.
To this end, we consider the $f(R)$ contribution \eqref{action-f(R)-fixed} plus additional terms involving 
Lagrange multipliers and gravitational Chern-Simons contributions.

Let us first recall the decomposition of the nonmetricity $Q_{\lambda\mu\nu}$ and torsion
${T_{\lambda\mu}}^{\nu}$ in a trace and traceless part. In three dimensions, one has \cite{Iosifidis:2019jgi}
\begin{align}
& Q_{\lambda\mu\nu} = \frac25 Q_\lambda g_{\mu\nu} - \frac15\tilde{Q}_\lambda g_{\mu\nu} +
\frac35 g_{\lambda(\nu}\tilde{Q}_{\mu)} - \frac15 g_{\lambda(\nu} Q_{\mu)} + \Omega_{\lambda\mu\nu}\,, 
\label{gennm} \\
& {T_{\lambda\mu}}^\nu = {\delta_{[\mu}}^{\nu} T_{\lambda]} + {S_{\lambda\mu}}^\nu\,, \label{gentor}
\end{align}
where the traces $Q_\lambda$ (the Weyl vector), $\tilde{Q}_\lambda$ and $T_\lambda$ were defined
in the previous section, while $\Omega_{\lambda\mu\nu}$ and ${S_{\lambda\mu}}^\nu$ denote
the traceless parts of the nonmetricity and torsion respectively.
In the following, we shall need the variation of the latter w.r.t.~the metric and the connection, which is
given by \cite{Iosifidis:2019jgi}
\begin{equation}
\delta_g Q_{\rho\alpha\beta} = \partial_\rho\left(g_{\mu\alpha} g_{\nu\beta} \delta g^{\mu\nu}\right) - 
2 g_{\lambda\mu} g_{\nu(\alpha} {\Gamma^\lambda}_{\beta)\rho}\delta g^{\mu\nu}\,, \qquad
\delta_g {T_{\mu\nu}}^\alpha = 0\,, \label{deltag-QT}
\end{equation}
and
\begin{equation}
\delta_\Gamma Q_{\rho\alpha\beta} = 2\delta^\nu_\rho\delta^\mu_{(\alpha} g_{\beta)\lambda}
\delta {\Gamma^\lambda}_{\mu\nu}\,, \qquad
\delta_\Gamma {T_{\alpha\beta}}^\lambda = \delta^{[\mu}_{\alpha}\delta^{\nu]}_{\beta}
\delta {\Gamma^\lambda}_{\mu\nu}\,. \label{deltaconn-QT}
\end{equation}
These imply
\begin{align}
& \delta_g Q_\rho = \partial_\rho\left(g_{\mu\nu}\delta g^{\mu\nu}\right)\,, \qquad
\delta_g T_\mu =0\,, \\
& \delta_g \tilde{Q}_\beta = \delta g^{\mu\nu} [g_{\nu\beta} g^{\rho\alpha}\partial_\rho g_{\mu\alpha}
+ {\Gamma^\lambda}_{\mu\nu} g_{\lambda\beta} - g^{\rho\sigma} {\Gamma^\alpha}_{\rho\sigma}
g_{\mu\alpha} g_{\nu\beta}] + g_{\nu\beta}\partial_\mu\delta g^{\mu\nu}\,, \label{deltag-traces}
\end{align}
and
\begin{equation}
\delta_\Gamma Q_\rho = 2\delta^\nu_\rho\delta^\mu_\lambda\delta {\Gamma^\lambda}_{\mu\nu}\,,
\qquad
\delta_\Gamma\tilde{Q}_\beta = (g^{\mu\nu} g_{\beta\lambda} + \delta^\mu_\beta\delta^\nu_\lambda) 
\delta {\Gamma^\lambda}_{\mu\nu}\,, \qquad
\delta_\Gamma T_\alpha = \delta^{[\mu}_{\alpha}\delta^{\nu]}_{\lambda}\delta
{\Gamma^\lambda}_{\mu\nu}\,. \label{deltaconn-traces}
\end{equation}
We propose the action
\begin{equation}\label{actiontotfin}
\begin{aligned}
S & = \frac1{2\kappa^2}\int d^3 x\left[\sqrt{-g} f(R) + \frac1{2\mu}\epsilon^{\mu\nu\rho} Q_\rho 
\hat{R}_{\nu\mu}\right] + \int d^3 x\epsilon^{\mu\nu\rho}\left[\chi_{\rho\mu}\left(-\frac13 Q_\nu + 
\tilde{Q}_\nu\right) + \zeta_{\nu\sigma} {T_{\rho\mu}}^\sigma\right] \\
& + \frac3{\kappa^2\mu}\int d^3 x\epsilon^{\mu\nu\rho}\left( {\Gamma^\sigma}_{\tau\rho}
\partial_\mu {\Gamma^\tau}_{\sigma\nu} + \frac23 {\Gamma^\tau}_{\sigma\mu}
{\Gamma^\sigma}_{\alpha\nu} {\Gamma^\alpha}_{\tau\rho}\right)\,,
\end{aligned}
\end{equation}
where $f(R)=R^{3/2}$, $\hat{R}_{\mu\nu} := {R^\lambda}_{\lambda\mu\nu} =
\partial_{[\mu}Q_{\nu]}$ denotes the homothetic curvature tensor, and $\mu$ is a Chern-Simons
coupling constant. Note that \eqref{actiontotfin} contains a Chern-Simons term both for the Weyl vector
and the connection $\Gamma$. In \eqref{actiontotfin} we also introduced the Levi-Civita symbol 
$\epsilon^{\mu\nu\rho}=\sqrt{-g}\varepsilon^{\mu\nu\rho}$, where $\varepsilon^{\mu\nu\rho}$ is the 
Levi-Civita tensor. $\chi_{\mu\nu}=-\chi_{\nu\mu}$ and $\zeta_{\mu\nu}$ are Lagrange
multipliers\footnote{The action \eqref{actiontotfin} is diffeomorphism-invariant by construction,
since it can be written as $S=\int d^3 x\sqrt{-g}\Psi$, where $\Psi$ transforms as a scalar under
general coordinate transformations.}. The idea that gravitational Chern-Simons terms may be
useful to find an action principle for the three-dimensional EW equations appeared for the first time
in \cite{pt,Cacciatori:2005wz}. In particular, the suggestion to use a CS term for the Weyl vector
occurs as a final comment in \cite{pt}.
\eqref{actiontotfin} can also be written in the form
\begin{equation}\label{actiontotfin1}
\begin{aligned}
S & = \frac1{2\kappa^2}\int d^3 x\left[\sqrt{-g} f(R) + \frac1{2\mu}\epsilon^{\mu\nu\rho} Q_\rho 
\hat{R}_{\nu\mu}\right] + \int d^3 x\epsilon^{\mu\nu\rho}\left[\chi_{\rho\mu}\left(- \frac13 Q_\nu + 
\tilde{Q}_\nu\right) + \zeta_{\nu\sigma} {T_{\rho\mu}}^\sigma\right] \\
& + \frac3{2\kappa^2\mu}\int d^3 x\epsilon^{\mu\nu\rho}\left( {R^\sigma}_{\tau\mu\nu} 
{\Gamma^\tau}_{\sigma\rho} - \frac23 {\Gamma^\tau}_{\sigma\mu} {\Gamma^\sigma}_{\alpha\nu} 
{\Gamma^\alpha}_{\tau\rho}\right)\,.
\end{aligned}
\end{equation}
From the variation of \eqref{actiontotfin} w.r.t.~$\chi_{\mu\nu}$ and $\zeta_{\mu\nu}$ we
get respectively
\begin{equation}\label{qt}
\tilde{Q}_\mu = \frac13 Q_\mu\,,
\end{equation}
\begin{equation}
\varepsilon^{\mu\rho\sigma} {T_{\rho\sigma}}^\nu = 0\,. \label{eq:dual-T=0}
\end{equation}
The latter implies vanishing torsion,
\begin{equation}\label{zerotor}
{T_{\rho\sigma}}^\nu = 0\,.
\end{equation}
Varying w.r.t~${\Gamma^\lambda}_{\mu\nu}$ and using \eqref{deltaconn-QT}, \eqref{deltaconn-traces},
one obtains
\begin{equation}\label{eq:ConnTorCS}
\begin{split}
& {P_\lambda}^{\mu\nu} + {\delta_\lambda}^\nu g^{\mu\sigma}\frac{\partial_\sigma f'}{f'} - g^{\mu\nu} 
\frac{\partial_\lambda f'}{f'} + \frac2{\mu f'}\varepsilon^{\nu\rho\sigma}(3 {R^\mu}_{\lambda\rho\sigma}
- {\delta_\lambda}^\mu \hat{R}_{\rho\sigma}) \\
& + \frac{2\kappa^2}{f'}\left({\delta_\lambda}^\nu\varepsilon^{\mu\rho\sigma}\chi_{\rho\sigma} -
\frac23 {\delta_\lambda}^\mu\varepsilon^{\nu\rho\sigma}\chi_{\rho\sigma} + g_{\lambda\tau} 
\varepsilon^{\tau\rho\sigma}\chi_{\rho\sigma} g^{\mu\nu} + \varepsilon^{\mu\nu\rho}
\zeta_{\rho\lambda}\right) = 0\,,
\end{split}
\end{equation}
where ${P_\lambda}^{\mu\nu}$ is the Palatini tensor defined in \eqref{palatini}.
In the following, we shall consistently set the traceless part of the nonmetricity to zero,
\begin{equation}\label{omzero}
\Omega_{\lambda\mu\nu} = 0\,.
\end{equation}
Notice that \eqref{qt} and \eqref{omzero} are precisely the constraints
needed to get the Weyl nonmetricity \eqref{eq:Weyl-nonmetr}: Plugging \eqref{qt} and \eqref{omzero}
into the decomposition \eqref{gennm} gives $Q_{\lambda\mu\nu}=g_{\mu\nu}Q_\lambda/3$,
and thus
\begin{equation}
\nabla_\lambda g_{\mu\nu} = -Q_{\lambda\mu\nu} = -\frac13 Q_\lambda g_{\mu\nu}\,,
\end{equation}
which coincides with \eqref{eq:Weyl-nonmetr} for $h_\lambda=-Q_\lambda/6$.

Using \eqref{qt}, \eqref{zerotor}, \eqref{omzero}, and writing the Palatini tensor explicitly in terms of the 
Weyl vector $Q_\mu$, \eqref{eq:ConnTorCS} becomes
\begin{equation}\label{eqconnstart0}
\begin{split}
& \frac16(g^{\mu\nu} Q_\lambda - {\delta_\lambda}^\nu Q^\mu) + {\delta_\lambda}^\nu 
\frac{\partial^\mu f'}{f'} - g^{\mu\nu}\frac{\partial_\lambda f'}{f'} + \frac6{\mu f'}\left[2
\varepsilon^{\alpha\nu\beta} g_{\lambda\alpha} {{\tilde R}^\mu}_{\,\,\,\beta} + \varepsilon^{\alpha\mu\nu} 
(g_{\alpha\lambda}\tilde{R} - 2{\tilde R}_{\alpha\lambda})\right] \\
& + \frac1{\mu f'}\left[\frac13\varepsilon^{\alpha\mu\nu}(g_{\alpha\lambda} Q_\rho Q^\rho -
Q_\alpha Q_\lambda + 6\tilde{\nabla}_\alpha Q_\lambda) + \varepsilon^{\alpha\nu\beta}
g_{\alpha\lambda}\left(\frac13 Q_\beta Q^\mu - 2\tilde{\nabla}_\beta Q^\mu\right)
\right] \\
& + \frac{2\kappa^2}{f'}\left({\delta_\lambda}^\nu\varepsilon^{\mu\rho\sigma}\chi_{\rho\sigma} -
\frac23 {\delta_\lambda}^\mu\varepsilon^{\nu\rho\sigma}\chi_{\rho\sigma} + g_{\lambda\tau} 
\varepsilon^{\tau\rho\sigma}\chi_{\rho\sigma} g^{\mu\nu} + \varepsilon^{\mu\nu\rho}
\zeta_{\rho\lambda} \right) = 0\,.
\end{split}
\end{equation}
Here, $\tilde\nabla$ denotes the Levi-Civita connection, and ${\tilde R}_{\mu\nu}$, $\tilde R$
are respectively its Ricci tensor and scalar curvature. We also used \eqref{gendecompaffconn},
\eqref{distortion}, \eqref{gennm} and \eqref{gentor} to express the Riemann tensor in terms of its
Levi-Civita part and the Weyl vector.

The $\lambda ,\mu$ trace of \eqref{eqconnstart0} leads to
\begin{equation}\label{zetasymm}
\frac{2\kappa^2}{f'}\varepsilon^{\nu\rho\sigma}\zeta_{\rho\sigma} = 0\quad\Rightarrow\quad
\zeta_{[\rho\sigma]} = 0\,.
\end{equation}
Taking this into account and considering the $\lambda ,\nu$ trace of \eqref{eqconnstart0}, we obtain
\begin{equation}\label{lambdanutrace}
\frac{10\kappa^2}{3f'}\varepsilon^{\mu\rho\sigma}\chi_{\rho\sigma} - \frac16 Q^\mu - \frac1{\mu f'} 
\varepsilon^{\mu\rho\sigma}\hat{R}_{\rho\sigma} + \frac{\partial^\mu f'}{f'} = 0\,.
\end{equation}
Plugging \eqref{zetasymm} and \eqref{lambdanutrace} into \eqref{eqconnstart0}, and taking the
$\mu,\nu$ trace of the resulting equation, one finds
\begin{equation}\label{chizero}
\chi_{\rho\sigma} = 0\,,
\end{equation}
and thus \eqref{lambdanutrace} reduces to
\begin{equation}\label{cdfp}
\frac{\partial^\mu f'}{f'} = \frac16 Q^\mu + \frac1{\mu f'}\varepsilon^{\mu\rho\sigma}
\hat{R}_{\rho\sigma}\,.
\end{equation}
With \eqref{chizero} and \eqref{cdfp}, \eqref{eqconnstart0} becomes
\begin{equation}\label{expr1}
\begin{split}
& ({\delta_\lambda}^\nu\varepsilon^{\mu\rho\sigma} - \varepsilon^{\alpha\rho\sigma}
g_{\alpha\lambda} g^{\mu\nu})\hat{R}_{\rho\sigma} + 6\left[ 2\varepsilon^{\alpha\nu\beta}
g_{\lambda\alpha} {\tilde{R}^\mu}_{\,\,\,\beta}
+ \varepsilon^{\alpha\mu\nu}(g_{\alpha\lambda}\tilde{R} - 2 {\tilde R}_{\lambda\alpha})\right] \\
& + \frac13\varepsilon^{\alpha\mu\nu}(g_{\alpha\lambda} Q_\rho Q^\rho - Q_\alpha Q_\lambda
+ 6\tilde{\nabla}_\alpha Q_\lambda) + \varepsilon^{\alpha\nu\beta} g_{\alpha\lambda}\left(\frac13 
Q_\beta Q^\mu - 2\tilde{\nabla}_\beta Q^\mu\right)
+ 2\kappa^2\mu\varepsilon^{\mu \nu \rho} \zeta_{\lambda \rho} = 0\,.
\end{split}
\end{equation}
Before proceeding with the analysis of \eqref{expr1}, consider the variation of \eqref{actiontotfin}
w.r.t.~$g^{\mu\nu}$. To this end, observe that the Chern-Simons terms are topological, i.e., independent
of the metric up to boundary terms. If we use moreover \eqref{omzero} and \eqref{chizero}, it is 
straightforward to shew that the equations of motion of $g^{\mu\nu}$ are just given by the Einstein-Weyl
equations \eqref{eq:EW} for $n=3$,
\begin{equation}
R_{(\mu\nu)} - \frac R3 g_{\mu\nu} = 0\,. \label{eq:EW3d}
\end{equation}
Notice that, if one kept $\Omega_{\lambda\mu\nu}\neq0$, there would be additional pieces in 
\eqref{eq:EW3d}, so that one would not obtain precisely the EW equations, but something more general
that contains a traceless part of the nonmetricity.
Recall that \eqref{eq:EW3d} is supplemented by \eqref{qt}, \eqref{zerotor} and \eqref{omzero},
and that we get a nonexact Weyl vector whose dynamics is governed by \eqref{cdfp}.
In terms of its Levi-Civita part and the Weyl vector, \eqref{eq:EW3d} reads
\begin{equation}\label{eq:EW3dexpl}
\tilde{R}_{\mu\nu} - \frac13 g_{\mu\nu}\tilde{R} - \frac1{108} g_{\mu\nu} Q_\rho Q^\rho + \frac1{36} 
Q_\mu Q_\nu + \frac1{18} g_{\mu\nu}\tilde{\nabla}_\rho Q^\rho - \frac16\tilde{\nabla}_{(\mu} Q_{\nu)}
= 0\,.
\end{equation}
Plugging this into \eqref{expr1}, the latter becomes
\begin{equation}\label{expr2}
\begin{split}
& {\delta_\lambda}^\nu\varepsilon^{\mu\rho\sigma}\hat{R}_{\rho\sigma} - \varepsilon^{\alpha\rho\sigma} 
g_{\alpha\lambda} g^{\mu\nu}\hat{R}_{\rho\sigma} - 2\varepsilon^{\alpha\mu\nu}\hat{R}_{\lambda\alpha} 
+ 2\varepsilon^{\alpha\nu\rho} g_{\alpha\lambda} {\hat{R}^\mu}_{\,\,\,\beta}\\
& +\varepsilon^{\alpha\mu\nu} g_{\alpha\lambda}\left(\frac19 Q_\rho Q^\rho - 2\tilde{R} + \frac43 
\tilde{\nabla}_\rho Q^\rho\right) + 2\kappa^2\mu\varepsilon^{\mu\nu\rho}\zeta_{\lambda\rho} = 0\,.
\end{split}
\end{equation}
\eqref{cdfp} is equivalent to
\begin{equation}\label{dual}
\hat{R}^{\rho\sigma} = \varepsilon^{\rho\sigma\tau}\Upsilon_\tau\,,
\end{equation}
where
\begin{equation}\label{upsilon}
\Upsilon_\tau := \frac12\mu\varepsilon^{\rho\sigma\tau}\left(\frac16 f' Q_\tau - \partial_\tau f'\right)\,.
\end{equation}
Using this in \eqref{expr2}, one easily shows that the terms involving the homothetic curvature tensor 
identically vanish, and we are left with
\begin{equation}\label{expr3}
\varepsilon^{\alpha\mu\nu} g_{\alpha\lambda}\left(\frac19 Q_\rho Q^\rho - 2\tilde{R} + \frac43 \tilde{\nabla}_\rho Q^\rho\right) + 2\kappa^2\mu\varepsilon^{\mu\nu\rho}\zeta_{\lambda\rho} = 0\,,
\end{equation}
or equivalently
\begin{equation}\label{zetaexpr}
\zeta_{\mu\nu} = \frac1{\kappa^2\mu} g_{\mu\nu}\left(\tilde{R} - \frac1{18} Q_\rho Q^\rho - \frac23 
\tilde{\nabla}_\rho Q^\rho\right) = \frac1{\kappa^2\mu} g_{\mu\nu} R\,,
\end{equation}
where we used \eqref{qt}, \eqref{zerotor} and \eqref{omzero} in the last step.
Summarizing, one has
\begin{equation}\label{nmform}
Q_{\lambda\mu\nu} = \frac13 Q_\lambda g_{\mu\nu}\,,
\end{equation}
which corresponds to Weyl nonmetricity, together with \eqref{zerotor}, \eqref{cdfp} and \eqref{eq:EW3d}.
The latter are precisely the Einstein-Weyl equations in three dimensions.
The final form of the connection, obtained by plugging \eqref{zerotor} and \eqref{nmform} into 
\eqref{distortion}, results to be
\begin{equation}\label{connwithWeylnm}
{\Gamma^\lambda}_{\mu\nu} = \tilde{\Gamma}^\lambda_{\phantom{\lambda}\mu\nu} -
\frac16 g_{\mu\nu} Q^\lambda + \frac13 {\delta_{(\mu}}^\lambda Q_{\nu)}\,.
\end{equation}
With $f(R)=R^{3/2}$, eq.~\eqref{cdfp} becomes
\begin{equation}\label{finaleq}
\partial_\mu\ln R = \frac13 Q_\mu + \frac4{3\mu\sqrt R} g_{\mu\tau}\varepsilon^{\rho\sigma\tau} 
\hat{R}_{\rho\sigma}\,.
\end{equation}
As we can see, the Weyl vector is not exact and possesses a nontrivial dynamics. In particular,
\eqref{finaleq} is a differential equation containing only the Weyl vector and the metric.
Dualizing \eqref{finaleq} gives the generalized monopole equation (cf.~\cite{Jones:1985pla}),
\begin{equation}
dh = \star(d\Sigma + h\Sigma)\,, \label{eq:gen-monop}
\end{equation}
with the one-form $h$ and the function $\Sigma$ respectively defined by
$h_\lambda=-Q_\lambda/6$ and $\Sigma=\mu\sqrt R/4$. Actually, \eqref{eq:gen-monop}
represents a special case of the generalized monopole equation, since the latter has the exterior
derivative of any one-form $\omega$ on the lhs. If $\Sigma$ were constant (this can always be
achieved by a Weyl rescaling \eqref{weylresc}, under which $\Sigma\mapsto e^{-\Omega}\Sigma$),
\eqref{eq:gen-monop} would boil down to $dh=\star h\Sigma$, which is the self-duality condition
(3) of \cite{Townsend:1983xs} in three dimensions. We can thus regard \eqref{eq:gen-monop}
as a conformally invariant generalization of the three-dimensional self-duality condition.

\section{Final remarks}\label{final}

A longstanding open mathematical problem has been the construction of an action principle
for the Einstein-Weyl equations.
In this paper, we presented for the first time such an action in three dimensions, given by a
metric affine $f(R)$ gravity contribution plus additional pieces involving Lagrange multipliers and 
gravitational Chern-Simons terms. To be more precise, our model contains, in addition to
the Weyl nonmetricity, also a traceless part. However, the latter can be consistently set to zero,
and in this case our equations of motion boil down to the EW equations.

Let us spend some words on the matter coupling of \eqref{actiontotfin}.
As mentioned in section \ref{review}, when one adds matter to metric affine $f(R)$ theories of gravity,
the projective invariance of the action imposes the constraint ${\Delta_\mu}^{\mu\nu}=0$
on the hypermomentum. However, if we couple matter to the theory developed in the present paper, we
see that the extra contributions we have introduced in the action break projective invariance.
Thus, in particular, ${\Delta_\mu}^{\mu\nu}\neq 0$, and the inconsistency mentioned above does not
appear anymore.

It remains to be seen if our results can be extended to higher dimensions. If so, this would probably
involve topological terms like e.g.~$BF$ actions, in addition to $f(R)$ gravity. Note
that Chern-Simons terms were considered in modifications of four-dimensional general
relativity \cite{Jackiw:2003pm, Alexander:2009tp, Hehl:1990ir}.
We hope to come back to this point in a future publication.

Finally, it would also be interesting to explore possible cosmological applications of \eqref{actiontotfin},
along the lines of refs.~\cite{Capozziello:2019wfi,Capozziello:2019qlt}, which are based on $f(R)$
gravity theories with a Levi-Civita connection. In this context, \cite{Iosifidis:2020gth} presented a model for 
cosmological hyperfluids, i.e., fluids with intrinsic hypermomentum that induce spacetime torsion and 
nonmetricity.

\section*{Acknowledgements}

This work was supported partly by INFN and by MIUR-PRIN contract 2017CC72MK003.
The authors would like to thank L.~Andrianopoli, R.~D'Auria
and M.~Trigiante for inspiring discussions.

\end{document}